%% file: charm2.tex
\documentclass[final,5p,times,twocolumn]{elsarticle}
 \biboptions{comma,sort&compress}

\usepackage{graphicx}
\usepackage{here}

\def\beq{\begin{equation}}
\def\eeq{\end{equation}}
\def\bea{\begin{eqnarray}}
\def\eea{\end{eqnarray}}
\def\bq{\begin{quote}}
\def\eq{\end{quote}}

\def\nnb{\nonumber}
\def\ga{\left(}
\def\dr{\right)}

\def\lrar{\Longrightarrow}
																
\def\nnb{\nonumber}
\def\la{\langle}
\def\ra{\rangle}
\def\nin{\noindent}
\def\ba{\vspace*{-0.2cm}\begin{array}}
\def\ea{\end{array}\vspace*{-0.2cm}}

\def\b{$\bullet~$}
\def\als{\alpha_s}

\def\gg2{ \la\alpha_s G^2 \ra}
\def\gg3{g^3f_{abc}\la G^aG^bG^c \ra}
\def\ggg4{\la\als^2G^4\ra}

\journal{Physics Letters B}

\begin{document}

\begin{frontmatter}

\title{Gluon condensates  and $c$, $b$ quark masses from quarkonia ratios of  moments} 

 \author[label1]{Stephan Narison\corref{cor1} }
   \address[label1]{Laboratoire
de Physique Th\'eorique et Astroparticules, CNRS-IN2P3,  
Case 070, Place Eug\`ene
Bataillon, 34095 - Montpellier Cedex 05, France.}
\cortext[cor1]{Corresponding author}
\ead{snarison@yahoo.fr}


\pagestyle{myheadings}
\markright{ }
\begin{abstract}
\noindent
We extract (for the first time)  the ratio of the gluon condensate $\la g^3f_{abc}G^3\ra/\la \alpha_s G^2\ra$  expressed in terms of the liquid instanton radius $\rho_c$ from charmonium moments sum rules by examining the effects of $\la \alpha_s G^2\ra$ in the determinations of both $\rho_c$ and  the running $\overline{\rm MS}$ mass $\overline{m}_c(m_c)$. Using
a global analysis of selected ratios of moments at different $Q^2=0,~4m_c^2$ and $8m_c^2$ and
keeping  $\la \alpha_s G^2\ra$ from 0.06 GeV$^4$, where the estimate of $\rho_c$ is almost independent of $\la \alpha_s G^2\ra$,
we deduce: $\rho_c=0.98(21)$ GeV$^{-1}$ corresponding to $ \la g^3f_{abc}G^3\ra= (31\pm 13){\rm GeV}^2\la \alpha_s G^2\ra $. The value of $\overline{m}_c(m_c)$ is less affected
(within the errors) by the variation of $\la \alpha_s G^2\ra$, where a common solution from different moments are reached for $ \la \alpha_s G^2\ra\geq$ 0.02 GeV$^4$. Using the values of $\la \alpha_s G^2\ra=0.06(2)$ GeV$^4$ from some other channels and the previous value of $\la g^3f_{abc}G^3\ra$, we deduce:  $\overline{m}_c(m_c)=1261(18)$ MeV and $\overline{m}_b(m_b)=4173(10)$ MeV, where an estimate of the 4-loops (${\cal O}(\alpha_s^3)$) contribution has been included. Our analysis  indicates that the errors in the determinations of the charm quark mass and of $\alpha_s$ without taking into account the ones of the  gluon condensates have been underestimated. 
 To that accuracy, one can deduce the running light and heavy quark masses and their ratios evaluated at $M_Z$, where it is remarkable to notice the approximate equalities: ${m}_{s}/m_u\approx {m}_{b}/m_s\approx  {m}_{t}/m_b\approx 51(4)$, which might reveal some eventual underlying novel symmetry of the quark mass matrix in some Grand Unified Theories.
 \end{abstract}

\end{frontmatter}
\section{Introduction}
\vspace*{-0.25cm}
 \nin
Non-zero values of the gluon condensates have been advocated by SVZ \cite{SVZ,ZAKA}. Indeed, the gluon condensates play an important r\^ole in gluodynamics (low-energy theorems,...) and in some bag models as they are directly related to the vacuum energy density (with standard notations):
\beq
E=-{\beta(\alpha_s)\over 8\alpha_s^2}\la \alpha_s G^2\ra~.
\eeq
Moreover, the gluon condensates enter in the OPE of the hadronic correlators \cite{SVZ} and then are important in the analysis of QCD spectral sum rules (QSSR), especially, in the heavy quarks and in the pure Yang-Mills  gluonia channels where the light quark loops and quark condensates\,$^1$\footnotetext[1]{The heavy quark condensate contribution can be absorbed into the gluon one through the relation \cite{SVZ}: 
$
\la \bar QQ\ra=-{\la\alpha_s G^2\ra/ (12\pi m_Q)}+...
$An analogous relation also occurs for the mixed quark-gluon condensate \cite{SNB1,SNB2,SNB3}.}
 are absent to leading order \cite{SNB1,SNB2,SNB3}. The SVZ value:
\beq
\la\alpha_s G^2\ra\simeq 0.04 ~{\rm GeV}^4~,
\eeq
extracted (for the first time) from charmonium sum rules \cite{SVZ} has been challenged by different
authors \cite{SNB1,SNB2,SNB3}. Though there are strong indications that the exact value of the gluon condensate is around this value or most likely 2 times this value as obtained from heavy quarks exponential moments \cite{BELL,SNB1,SNB2,SNB3}, heavy quark mass-splittings \cite{SNHeavy} and $e^+e^-$ \cite{LNT,PEROTTET,SNI}, most recent determinations 
from $\tau$-decay \cite{ALEPH,OPAL,DAVIER} (see however \cite{SNTAU}) and the previous charmonium moments \cite{IOFFE} indicate that its value is not well determined. 
In fact, at present, the structure of the QCD vacuum is not yet under a good control.  If one follows the SVZ idea based on the ordinary OPE, the QCD confinement can be parametrized by the sum of quark and gluon condensates of higher and higher dimensions\,\footnote{A possible existence of an additional $1/Q^2$ term induced by large order terms of PT series has been discussed  in \cite{CNZ,SNREV}.}.
In order to estimate the higher dimension condensates, one usually assumes factorization using vacuum
saturation (leading $1/N_c$ approximation). However, in many examples, this assumption appears to be badly violated \cite{LNT,PEROTTET,ALEPH,OPAL,DAVIER,SNZ1,RAF04,SNVA,BORDES,MENES,DOMI,DOSCH}.
Different phenomenological works have been performed for understanding the complex structure of the QCD vacuum in the $V+A$ and $V-A$ channels of the light  flavours \cite{LNT,PEROTTET,ALEPH,OPAL,DAVIER,SNZ1,RAF04,SNVA,BORDES,MENES,DOMI,DOSCH} and from lattice calculations \cite{GIACO,GIACO2,RAKOW}. Here, we  shall estimate (for the first time) the ratio of the dimension-6 $\la g^3f_{abc}G^3\ra$ over the dimension-4 $\la \alpha_s G^2\ra$ gluon condensates using charmonium sum rules\,\footnote{The $\la g^3f_{abc}G^3\ra$ condensate does not contribute in the chiral limit $m_q=0$ in the vector and axial-vector channels of light flavours.}, in the aim to clarify the different inaccurate proposals from some instanton liquid models. In so doing, we find that it is convenient to introduce the instanton radius $\rho_c$:
\beq
{\la g^3f_{abc}G^3\ra\over \la \alpha_s G^2\ra}= {4\over 5}{12\pi\over\rho_c^2}~.
\label{eq:cond}
\eeq
The value of $\rho_c$ ranges from 1/3 fm=1.5 GeV$^{-1}$ \cite{SHURYAK}, 0.5 fm=2.5 GeV$^{-1}$ \cite{IOFFE2}  to 0.9 fm= 4.5 GeV$^{-1}$ \cite{SVZ}. As $\la g^3f_{abc}G^3\ra$ contributes like 1/$\rho_c^2$ in the OPE analysis, a more precise value of $\rho_c$ is crucial for checking the convergence of the OPE. The estimate of $\rho_c$ from charmonium sum rules is feasible as the light quark condensates $m_q\la \bar qq\ra$ contributes, to higher loop order and which are chiral suppressed are negligible, while the heavy quark condensate contribution can be absorbed into the gluon one as mentioned earlier.  
\section{Moment sum rules }
\label{sec:qssr}
\vspace*{-0.25cm}
 \nin
 Here, we shall be concerned with the two-point correlator of a heavy quark $Q$:
 \bea
&& -\ga g^{\mu\nu}q^2-q^\mu q^\nu\dr \Pi_Q(q^2)\equiv \nnb\\
&& i\int d^4x ~e^{\rm-iqx}\la 0\vert {\cal T} J^\mu_Q(x)\ga J^\nu_Q(0)\dr^\dagger \vert 0\ra~,
 \eea
 where : $J_Q^\mu=\bar Q \gamma^\mu Q$ is the heavy quark neutral vector current. Im $\Pi_c(s)$ can be related to the charmonium leptonic widths and masses. In a narrow width approximation (NWA):
 \bea
 {\cal R}_c(t)&\equiv& 4\pi{\rm Im} \Pi_c(t+i\epsilon)\nnb\\
 &=&{N_c\over Q_c^2 \alpha^2}\sum_{}M_{\psi}\Gamma_{\psi\to e^+e^-}\delta(\ga t-M^2_{\psi}\dr~,
 \eea
 where $N_c=3$; $M_{\psi}$ and $\Gamma_{\psi\to e^+e^-}$ are the mass and leptonic width of the $J/\psi$ mesons; $Q_c=2/3$ is the charm electric charge in units of $e$; $\alpha=1/133.6$ is the running electromagnetic coupling evaluated at $M^2_{\psi}$. We shall use the experimental values of the $J/\psi$ parameters compiled in Table \ref{tab:psi}.\\
 \vspace*{-.5cm}
{\scriptsize
\begin{table}[hbt]
\setlength{\tabcolsep}{1.5pc}
 \caption{\scriptsize    Masses and electronic widths of the  $J/\psi$ family from PDG 08 \cite{PDG}. }
{\small
\begin{tabular}{lll}
&\\
\hline
Name&Mass [MeV]&$\Gamma_{J/\psi\to e^+e^-}$ [keV] \\
\hline
\\
$J/\psi(1S)$&3096.916(11)&5.55(14)\\
$\psi(2S)$&3686.093(34)&2.33(7)\\
$\psi(3770)$&3775.2(1.7)&0.259(16)\\
$\psi(4040)$&4039(1)&0.86(7)\\
$\psi(4160)$&4153(3)&0.83(7)\\
$\psi(4415)$&4421(4)&0.58(7)\\
\\
\hline
\end{tabular}
}
\label{tab:psi}
\end{table}
}
\nin
\\
 Different forms of QSSR exist in the literature \cite{SNB1,SNB2,SNB3}. We shall work here with the moments:
 \beq
 {\cal M}_n\ga -q^2\equiv Q^2\dr=\int_{4m_Q^2}^\infty dt {{\cal R}_c(t,m_c^2)\over (t+Q^2)^{n+1}}~,
 \eeq
and more likely with their ratios:
 \beq
 r_{n/n+1}(Q^2)={{\cal M}_n\over {\cal M}_{n+1}},~~~~r_{n/n+2}(Q^2)={{\cal M}_n\over {\cal M}_{n+2}}~,
 \eeq
 where the experimental sides are more precise than the absolute moments ${\cal M}_n$. Also, in the ratios, partial cancellations of different perturbative as well as non-perturbative terms occur, which render the QCD approximation more precise than in the absolute moments.\\
 The QCD sides of the sum rules are known in the literature since the original works of SVZ \cite{SVZ}. Their expressions at the subtraction scale $\nu^2=m_Q^2$ are given explicitly numerically in the Appendix of \cite{IOFFE} to 3-loops (${\cal O}(\alpha_s^2)$) accuracy in terms of the running heavy quark mass\,\footnote{We shall use these expressions in our analysis and we shall correct our final results on the quark masses by adding an estimate of the 4-loops (${\cal O}(\alpha_s^3)$) contributions.} using the pQCD results of \cite{KUHN}, while  the $Q^2=0$ moments to 4-loops (${\cal O}(\alpha_s^3)$) are given in \cite{KUHN1} using the pQCD results in \cite{KUHN2,BOUG} \footnote{Some Pade approximants are given in \cite{MATEU}.}. Among the different moments
given in \cite{IOFFE}, we shall select three moments where both the $(\alpha_s)^n~(n=1,~2)$, the gluon condensate contributions and the effects of the higher resonances plus the QCD continuum are relatively small but not negligible. These conditions can be simultaneously satisfied by the moments \footnote{However, one should note that the accuracy of the $Q^2=0$ moments is less than that of the $Q^2\not=0$ moments, while one cannot use higher moments due to the bad convergence of the OPE in this case.}:
\bea
&& {\cal M}_{2, 3, 4}~~~ ~~~{\rm for}~~~~Q^2=0~,\nnb\\ 
&& {\cal M}_{8, 9, 10}~~~~~ {\rm for} ~~~~Q^2=4m_Q^2~,\nnb \\
 &&{\cal M}_{13, 14, 15}~~~ {\rm for} ~~~~Q^2=8m_Q^2~.
 \eea
One may also work with more moments but these will not bring newer informations. Lower moments are more sensitive to the experimental errors and to the QCD continuum while higher moments are more sensitive to higher dimension condensates which are not under a good control\,\footnote{The contributions of the dimension-8 condensates have been evaluated in \cite{RADYUSH} and can be sizeable if one assumes factorization which might not be applied here \cite{BAGAN}.}. Moreover, one can also note from the QCD expressions given by \cite{IOFFE} that for $n$ larger than in our previous selected choice, the signs of the pQCD corrections start to change compared to the original ones of the two-point correlator. A such change may introduce some systematical difficulties inherent to the approach for controlling the size of higher order terms\,\footnote{In \cite{PIVO},  some low energy gluon contributions to order $\alpha_s^3$ to the correlator
can invalidate the uses of the $Q^2=0$-moments for $n>4$ (see however \cite{SIGNER}).}.
{\scriptsize
\begin{table}[hbt]
\setlength{\tabcolsep}{0.9pc}
 \caption{\scriptsize    QCD expressions of the moments and their ratios normalized to the lowest order terms and given to order $\alpha_s^2$ and including the dimension-4 and -6 gluon condensates derived from  \cite{IOFFE}; $d_4\equiv  {\la \alpha_sG^2\ra/ (4m_c^2)^2}\simeq (1.49\pm 0.50)\times 10^{-3}$ if we use $m_c=1.261$ GeV from Table \ref{tab:res} and $\la \alpha_sG^2\ra=0.06$ GeV$^4$ in Eq. \ref{eq:g2}; $\rho_{64}\equiv\la g^3f_{abc} G^3\ra/(\la\alpha_s G^2\ra 4m_c^2)\simeq (4.88\pm 2.05)$ is the ratio between the dimension-6 and -4 gluon condensate contributions if we use the numerical value in Eq.~\ref{eq:gg3}; $a_s\equiv \alpha_s/\pi$.  }
{\small
\begin{tabular}{ll}
&\\
\hline
Mom&QCD expression \\
\hline
\\
$Q^2$=0\\
${\cal M}_2$&$1+2.427a_s+6.110a_s^2-d_4\ga18.61-0.83{\rho_{64}}\dr$\\
${\cal M}_3$&$1+1.917a_s+6.115a_s^2-d_4\ga45.71-4.00{\rho_{64}}\dr$\\
${\cal M}_4$&$1+1.100a_s+4.402a_s^2-d_4\ga90.21-12.76{\rho_{64}}\dr$\\
$r_{2/3}$&$1+0.510a_s-0.983a_s^2+d_4\ga27.10-3.17{\rho_{64}}\dr$\\
$r_{2/4}^{1/2}$&$1+0.664a_s-0.095a_s^2+d_4\ga35.80-5.97{\rho_{64}}\dr$\\
\\
$Q^2=4m_c^2$\\
${\cal M}_8$&$1+1.118a_s+4.253a_s^2-d_4\ga77.51-5.02{\rho_{64}}\dr$\\
${\cal M}_9$&$1+0.601a_s+2.700a_s^2-d_4\ga104.86-9.20{\rho_{64}}\dr$\\
${\cal M}_{10}$&$1+0.045a_s+1.136a_s^2-d_4\ga137.83-15.63{\rho_{64}}\dr$\\
$r_{8/9}$&$1+0.517a_s+1.242a_s^2+d_4\ga27.35-4.18{\rho_{64}}\dr$\\
$r_{8/10}^{1/2}$&$1+0.537a_s+1.390a_s^2+d_4\ga30.16-5.31{\rho_{64}}\dr$\\
\\
$Q^2=8m_c^2$\\
${\cal M}_{13}$&$1+0.776a_s+3.061a_s^2-d_4\ga90.37-5.19{\rho_{64}}\dr$\\
${\cal M}_{14}$&$1+0.412a_s+1.909a_s^2-d_4\ga109.75-7.89{\rho_{64}}\dr$\\
${\cal M}_{15}$&$1+0.031a_s+0.770a_s^2-d_4\ga137.72-11.53{\rho_{64}}\dr$\\
$r_{13/14}$&$1+0.364a_s+1.002a_s^2+d_4\ga19.38-2.70{\rho_{64}}\dr$\\
$r_{13/15}^{1/2}$&$1+0.373a_s+1.065a_s^2+d_4\ga23.68-3.17{\rho_{64}}\dr$\\
\\
\hline
\end{tabular}
}
\label{tab:ope}
\end{table}
}
\nin
\\
We shall work with the ratios of moments:
\bea
&&r_{2/3}~~~~~{\rm and}~~~~~  r_{2/4}~~~~~~{\rm for} ~~~~~~Q^2=0\nnb\\
&&r_{8/9}~~~~~{\rm and} ~~~~~ r_{8/10}~~~~~{\rm for}~~~~~~Q^2=4m_c^2\nnb\\
&&r_{13/14}~~~{\rm and} ~~~~~ r_{13/15}~~~{\rm for}~~~~~~Q^2=8m_c^2
\label{eq:ratiomom}
\eea
and use as inputs, in this first step:
\beq
\overline{m}_c(m_c)=1.26(3)~{\rm GeV}~,
\label{eq:mcinput}
\eeq
as given by  different approaches using charmonium moments sum rules \cite{SVZ,SNB1,SNB2,SNB3,IOFFE,KUHN1,RRY,SNmc,JAMIN,PDG} 
and which we shall re-estimate later on. We shall also use:
\beq
\alpha_s(M_\tau)=0.3249(80) \lrar \alpha_s(m_c)\vert_{n_f=4}=0.408(14)
\label{eq:alphas}
\eeq
from $\tau$-decay \cite{SNTAU}; a value which agrees with the central value of the world
average \cite{PDG,BETHKE} when runned until $M_Z$. \\ The QCD expressions of the moments 
and their ratios are given in Table~\ref{tab:ope}.

\section{\boldmath$\rho_c$  from charmonium ratios of moments}
\label{sec:qssr2}
\vspace*{-0.25cm}
 \nin
 We shall work with the ratios of moments in Eq. \ref{eq:ratiomom}. 
 We parametrize the spectral function by a sum of the six $J/\psi$-like narrow resonances below 4.6 GeV\,\footnote{One can improve this parametrization by taking into account finite width corrections using BES data \cite{BES}, but these corrections will be negligible in the moments which we shall use.} and use its pQCD expression from $\sqrt{t}=(4.6\pm 0.1)$ GeV. We extract the value of $\rho_c$ for a large range of $\la\alpha_s G^2\ra$. 
\begin{figure}[hbt]
\begin{center}
\includegraphics[width=8.cm]{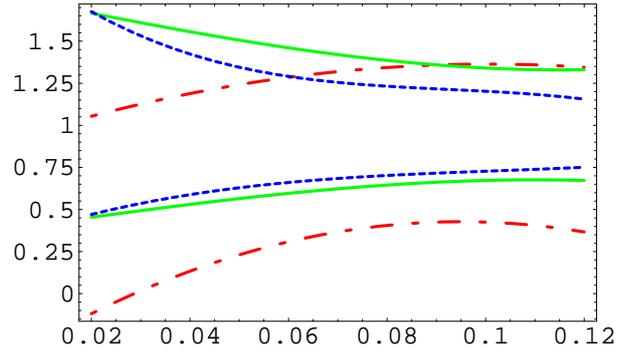}
\caption{\footnotesize Determinations of $\rho_c$ in GeV$^{-1}$ versus $\la\alpha_s G^2\ra$ in GeV$^4$ from different sets of moments :  $Q^2=0$ (red: dashed-dotted) ; $Q^2=4m_c^2$ (green: continuous); 
$Q^2=8m_c^2$ (blue: dotted).} 
\label{fig:rhoc}
\end{center}
\end{figure} 
{\scriptsize
\begin{table}[hbt]
\setlength{\tabcolsep}{1.5pc}
 \caption{\scriptsize    Ratio $\rho_c$ of the $\la g^3f_{abc}G^3\ra/\la \alpha_s G^2\ra$ as defined in Eq. \ref{eq:cond}, and value of $\overline{m}_c(m_c)$ from charmonium moments known to 3-loops. The errors in $\rho_c$ 
 come  from the choice of moments and from the error on $m_c$ given in Eq. \ref{eq:mcinput}. The ones due to $\alpha_s$, $\la \alpha_s G^2\ra$  and to the data on the $J/\psi$ family are negligible. The value of $m_c$ is  taken at $\la \alpha_s G^2\ra=0.06(2)$ GeV$^4$ as given in Eq. \ref{eq:g2}. The errors on $m_c$ come respectively from the ones of $\rho_c$ and $\la \alpha_s G^2\ra$. The ones due to $\alpha_s$ and the data are about 1 MeV each which are negligible. }
{\small
\begin{tabular}{lll}
&\\
\hline
Mom&$\rho_c$ [GeV$^{-1}$]&$m_c(m_c)$ [MeV] \\
\hline
\\
$Q^2$=0:\\
$r_{2/3},~r_{2/4}$&0.800(490)&1234(34)(8)\\
\\
$Q^2$=$4m_c^2:$ \\
$~r_{8/9},~r_{8/10}$&1.025(425)&1265(29)(9)\\
\\
$Q^2$=$8m_c^2:$\\
$r_{13/14},~r_{13/15}$&1.025(275)&1268(18)(7)\\
\\
\hline
Average & 0.98(21)&1261(15)
\\
\hline
\end{tabular}
}
\label{tab:res}
\end{table}
}
\nin
\\
One can see in Fig. \ref{fig:rhoc} that the results are very stable for all moments for $\la\alpha_s G^2\ra\geq 0.06$ GeV$^4$, from which we deduce the value of $\rho_c$ given in Table \ref{tab:res}. 
The errors in $\rho_c$ come respectively  from the values of $m_c$ and of the choice of the moments at given $Q^2$. These errors are included in the regions given in Fig. \ref{fig:rhoc}. 
 The one due  to $\alpha_s$ and to experiments are negligible . 
 Our final averaged result is:
 \beq
 \rho_c=0.98(21)~{\rm GeV}^{-1} ~~~{\rm for}~~~\la\alpha_s G^2\ra\geq 0.06~{\rm GeV}^4~,
 \label{eq:rho}
 \eeq
 which we consider as an improvement of the different estimates based on instanton liquid models \cite{SVZ,SHURYAK,IOFFE2} recalled in the introduction. However, our result agrees within the error with the one in \cite{SHURYAK}
 but is smaller by a factor of about 4 than the SVZ estimate \cite{SVZ}. Using Eq. \ref{eq:cond}, our result for $\rho_c$ corresponds to\,\footnote{A recent estimate using exponential sum rules and  including the dimension-eight condensate leads to a lower value $ \la g^3f_{abc}G^3\ra =(8.3\pm 1.0)~{\rm GeV}^2\la \alpha_s G^2\ra$ \cite{SNexp}, which is still higher than the previous dilute gas instanton estimates.}:
 \beq
 \la g^3f_{abc}G^3\ra =(31\pm 13)~{\rm GeV}^2\la \alpha_s G^2\ra~,
 \label{eq:gg3}
 \eeq
 indicating that it is much bigger than usually assumed in the literature. It is also smaller than  the  lattice result in $SU(2)$ pure Yang-Mills \cite{GIACO} and than a rough estimate extended to $SU(3)$ with dynamical fermions \cite{GIACO3}
 using the result in \cite{GIACO,GIACO2}. One should notice that this value
 of the gluon condensate has been extracted by assuming that the OPE including the dimension-6 gluon condensate gives a good description of the experimental data which implicitly assumes that the contributions of higher dimension condensates are negligible in the analysis. In other words, one may also interpret this value as the one of an {\it ``effective gluon condensate"} which parametrizes all higher dimension condensates contributing to the OPE.  
Fitted values of the higher dimensions vacuum condensates in the light quark channels have been also found to be larger than the vacuum saturation assumptions \cite{LNT,PEROTTET,SNZ1,RAF04,SNVA,DOSCH} and our results seem to go towards this direction. However, as we shall explicitly discuss later, these values of the gluon condensates remain still a correction compared to the one of the lowest order perturbative contribution in the ratios of moments which we use here and do not break the OPE.   
\section{\boldmath $\overline{m}_c(m_c)$ from charmonium ratios of moments}
\label{sec:qssr22}
\vspace*{-0.25cm}
 \nin

\begin{figure}[hbt]
\begin{center}
\includegraphics[width=8.cm]{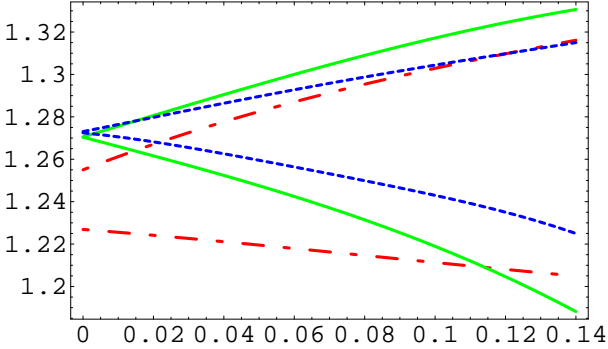}
\caption{\footnotesize  Determinations of $\overline{m}_c(m_c)$ in GeV versus $\la\alpha_s G^2\ra$ in GeV$^4$ from different sets of moments :  $Q^2=0$ (red: dashed-dotted) ; $Q^2=4m_c^2$ (green: continuous); 
$Q^2=8m_c^2$ (blue: dotted).}
\label{fig:mc}
\end{center}
\end{figure} 
\nin
 For extracting $\overline{m}_c(m_c)$, we equate the QCD and experimental sides of the ratios of moments $ r_{n/n+1}(Q^2)$ and  $r_{n/n+2} (Q^2)$ and solve exactly numerically these equations using 
 the Mathematica subroutine FindRoot. Though the equations look simple for the $Q^2$=0 - moment:
 \beq
 F_{th}(x)\equiv ax+b+{c\over x}+{d\over x^2}+{e\over x^3}=F_{exp}(x)
 \eeq
 $x\equiv m_c^2$, they are highly non-trivial for $Q^2=4m_c^2, 8m_c^2$ and for higher values of $n$ due to the appearance of the term:
 \beq
F_{exp}(x)\sim \sum_{\psi} {g^2_\psi\over (M^2_\psi+Q^2)^{n+1}}~,
 \eeq
  in the experimental side of the ratio of moments. 
 We show in Fig. \ref{fig:mc} the different solutions of $\overline{m}_c(m_c)$ versus $\la\alpha_s G^2\ra$ for each ratio $ r_{n/n+1}(Q^2)$ and  $r_{n/n+2} (Q^2)$ of $Q^2$-moments.  
 One can notice that common solutions of different ratios moments occur in the range of values:
 \beq
 \la\alpha_s G^2\ra \geq 0.02~{\rm GeV}^4~,
\label{eq:condensate}
\eeq
indicating that the central values of $\overline{m}_c(m_c)$ are not very sensitive to the one of $\la\alpha_s G^2\ra$. This feature confirms the unconclusive range of values obtained for $\la\alpha_s G^2\ra$ in \cite{IOFFE}. However, zero and negative values of $\la\alpha_s G^2\ra$ as obtained from some analysis of $\tau$-decays \cite{DAVIER}\,\footnote{One should notice that due to the kinematical structure of the original $\tau$-decay width \cite{BNP}, the gluon condensate contribution acquiers there an extra $\alpha_s$ coefficient compared with the one of the two-point correlator which suppresses its contribution and can render inaccurate its extraction from this observable.} are excluded by our present result and by the one in \cite{IOFFE}. The slight difference between \cite{IOFFE} with our analysis is that we put implicitly the previously determined value of $\la g^3f_{abc}G^3\ra$ and its correlation with $\la\alpha_s G^2\ra$ in the extraction of $\overline{m}_c(m_c)$. This fact explains the increase of the errors for increasing values of $\la\alpha_s G^2\ra$ in our analysis.  
In order to improve the determination of $m_c$, we use the values:
\beq
 \la\alpha_s G^2\ra= 6(2)\times 10^{-2} ~{\rm GeV}^4~,
 \label{eq:g2}
\eeq
obtained by enlarging the error of the average value 0.06(1) GeV$^4$ from the heavy quarkonia mass-splittings~\cite{SNHeavy}:
\beq
\la\alpha_s G^2\ra= 7.5(2.5)\times 10^{-2} ~{\rm GeV}^4~,
\eeq
 and from $e^+e^- \to I=1$ hadrons sum rules \cite{SNI}:
 \beq
\la\alpha_s G^2\ra= 6.1(0.7)\times 10^{-2} ~{\rm GeV}^4~.
\eeq
These previous values agree with the one about 0.069 GeV$^4$ obtained from $SU(3)$ lattice with dynamical fermions \cite{GIACO2}.
Using this value, we show in Table \ref{tab:res} the value of $\overline{m}_c(m_c)$ from each sets of ratios of moments from which we can deduce the mean value:
\beq
\overline{m}_c(m_c)\vert_{3-loops}=1261(15)~{\rm MeV}~,
\label{eq:mc3loop1}
\eeq
obtained from a 3-loop (${\cal O}(\alpha_s^2)$) expression of the ratios of moments. This value is more weighted by the one from the $Q^2=8m_c^2$ ratios of moments which give the most precise predictions.

\section{Comments on the results}
\vspace*{-0.25cm}
 \nin

{\it \b $Q^2=0$ moments}	\\
We note that our result $\overline{m}_c(m_c)=1234(35)$ MeV from the $Q^2=0$ moments  agrees within the error with the four-loops recent estimate 1279(13) MeV in \cite{KUHN1} based on the lowest $n=1$ moment. Our result is less precise due mainly to the errors induced by the presence of $\la g^3f_{abc}G^3\ra$ and of its correlated $\la\alpha_s G^2\ra$ condensate in our analysis and to the error induced by the choice of the ratio of moments as can be seen in Fig. \ref{fig:mc}. 
It is informative to compare the size of each QCD corrections in the OPE, that can be deduced from the QCD expressions given in Table \ref{tab:ope}:\\
-- One should first notice that each perturbative and non-perturbative corrections tends to partially cancel out in the ratios of moments, which render the QCD PT series and OPE more convergent for the ratios than for the corresponding individual moments. Therefore, one expects that these ratios of moments  can lead to more robust predictions. \\
-- Using the previous values of the QCD parameters and for definiteness $\la\alpha_s G^2\ra=0.06$ GeV$^4$, one finds that the contribution of $\la\alpha_s G^2\ra$ in ${\cal M}_2(0)$ is about -2.8\% which is
comparable with the one -1.6\% from $-0.23\alpha_s^3$ \cite{KUHN1}, while, for ${\cal M}_3(0)$, it is -6.8\% which is about the one 10\% from $\alpha_s^2$.
For ${\cal M}_2(0)$ the $\la g^3f_{abc}G^3\ra$ contribution is about 0.6\% which is about 1/3 of the $\alpha_s^3$ one, while for ${\cal M}_3(0)$ it is about 3\% compared with 2\% from  0.299$\alpha_s^3$ and with 10\% from $\alpha_s^2$. These features indicate that the non-perturbative corrections can be comparable with the PT radiative corrections and cannot be neglected like usually done in the exisiting literature (see e.g. \cite{KUHN1} and references therein). The same remark also applies to the extraction of $\alpha_s$ in \cite{STEIN0} from low-$n$ moments.\\
-- Finally, the leading experimental error due to the $J/\psi$ leptonic widths, which gives a strong limitation to the accuracy of the low $n$ moments (see e.g. \cite{KUHN1} and references therein), partially cancel out in the ratio of moments such that the experimental error  induces only a negligible error of about 1 MeV in the dermination of $m_c$. In the same way, the high mass states contributions to the spectral functions
are more suppressed in the ratios of moments, which then avoid some difficulties induced by the present data in the high-energy regions.  \\ 
\\
{\it \b $Q^2\not=0$ moments} \\
With these moments, we can work at larger values of $n$, where the experimental sides of the sum rules
become more accurate due to the increase of the weight of the lower mass resonances contributions in these moments:\\
--  Comparing  e.g. the QCD sides of the $n=2,Q^2=0$ ${\cal M}_2(0)$ with that of the $n=8, Q^2=4m_c^2~ {\cal M}_8(4m_c^2)$-moments, we find from Table \ref{tab:ope} that the sum of the PT corrections up to order $\alpha_s^2$ for ${\cal M}_8(4m_c^2)$ (21\%) are about 1/2 of the ones  for ${\cal M}_2(0)$ (42\%). Moreover, though the size of the sum of the NP terms increases from -2.2\% for ${\cal M}_2(0)$ to -9.6\% for ${\cal M}_8(4m_c^2)$, one can see that the ratio between the $\la g^3f_{abc}G^3\ra$ over the $ \la\alpha_s G^2\ra$ contributions are almost unchanged of about 20\%, indicating the good convergence of the OPE in the analysis. \\
-- One can also deduce from Table \ref{tab:ope}, that the PT and NP QCD corrections  are much lower for the ratios of moments. The sum of PT corrections is typically 6.5\% while the NP ones are  2\%. A convergence of the OPE is still observed though  the ratio between the $\la g^3f_{abc}G^3\ra$ over the $ \la\alpha_s G^2\ra$ contributions is  larger ($0.6\sim 0.8$) for the ratios of moments than   for the moments ($0.3\sim 0.4$). 
\\
-- Our results from these $Q^2\not=0$ ratios of moments given in Table \ref{tab:res} agree with the ones 1275(15) MeV obtained in \cite{IOFFE} though less accurate due to the effect of the error on $\la g^3f_{abc}G^3\ra$ included in our analysis.\\
\\
{\it \b Concluding remarks}\\
-- These previous facts indicate that 
a precise determination of $\overline{m}_c(m_c)$ and of $\alpha_s$ requires the inclusion of the non perturbative condensates which can induce large errors even for the lowest $Q^2=0$ moments and which have not been taken properly into account in the existing literature.\\
-- One can also note from Table \ref{tab:res} that the best estimate of $\overline{m}_c(m_c)$ comes from the $Q^2\not=0$ ratio of moments where the error due to the choice of the ratio of moments is smaller than in the case $Q^2=0$. 
\\
\\
{\it \b Error due to the subtraction point $\nu$}\\
Our previous results in Eq. \ref{eq:mc3loop1} have been obtained at the subtraction point $\nu^2=m_c^2$. If the whole series is known, the results should be independent of $\nu$. 
The knowledge of the PT series to 4-loops decreases the sensitivity of the results on $\nu$. 
To order $\alpha_s^2$ where the
moments have been evaluated, one can introduce this $\nu$-dependence through the replacement (see e.g.: \cite{SNB3,SNB1}):
\bea
\alpha_s(m_c)&\to& \alpha_s(\nu)\times \ga 1-{\beta_1}~{\alpha_s(\nu)\over\pi}\log{\nu\over  m_c}\dr~,
\label{eq:sub}
\eea
where $\beta_1=-(1/2)(11-2n_f/3)$ for $n_f$-flavours.
Taking
$
 0.5\leq \nu^2/m_c^2\leq 2~,
$
 we deduce from the $Q^2=8m_c^2$ ratio of moments:
 \beq
 \delta_{m_c}\vert_{\nu}=\pm 6~{\rm MeV}~,
 \eeq
 which we consider to be more conservative than the one of about 2-3 MeV given in \cite{IOFFE}. Alternatively, one can also minimize the $\nu$-dependence by working at large $Q^2$ and with low $n$ moments and after running down the result to $m_c$. In this way, one would obtain a slightly smaller error of about 3-5 MeV \cite{KUHN1}. \\
 \\
 {\it \b Shift due to Coulombic corrections}\\
The contribution due to
Coulombic corrections are expected to be negligible (about 1-2 MeV \cite{IOFFE}) because the system is still relativistic. In fact, the Coulomb radius:
 \beq
 r_{Coul}\approx {2\over m_cC_F\alpha_s(m_c)}\simeq 3 ~{\rm GeV}^{-1}~,
 \label{eq:coulomb}
 \eeq
($C_F=4/3$) is much larger than the confinement radius $r_{conf}\approx$ 1 GeV$^{-1}$. 
These corrections can even
 be made much smaller by working with a $Q^2\not=0$ moments rather than with a $Q^2=0$
 one, as the quark velocity behaves for large $n$ as:
 \beq
 v\approx \sqrt{\ga 1+{Q^2/ 4m_c^2}\dr/ n}~,
 \label{eq:velocity}
 \eeq
 which, e.g., for $Q^2= 8m_c^2$ and $n=14$, is about 0.46. This value is not small and not inside the nonrelativistic region. 
 We can approximately estimate this effect by working with the resummed Coulombic expression of the spectral function \cite{EICHTEN}:
 \beq
 {\cal R}_c\vert_{Coul}\simeq {3\over 2}v{x\over 1-e^{-x}}~,
  \label{eq:coulomb2}
 \eeq
 where: $x\equiv \pi C_F\alpha_s/v$, $C_F=4/3$ and $v=\sqrt{1-4m_Q^2/t}$. We compare the value of the ratio of moments using this perturbative expression for the spectral function with the one obtained from PT theory including radiative corrections. In the case $Q^2= 8m_c^2$ and $n=14$, where the most precise result is obtained, the corrections induced by the Coulombic contributions to the value of $m_c$ are negligible\,\footnote{Some further arguments justifying the smallness of these contributions can be found in \cite{IOFFE}.} :
\beq
 \delta_{m_c}\vert_{Coul}=-(0.4\pm 0.4)~{\rm MeV}~,
 \eeq 
 where we have assumed that our determination is known within 100\% error.
 
\section{$\overline{m}_c(m_c)$ to order ${\cal O}(\alpha_s^3)$}
\vspace*{-0.25cm}
 \nin
 {\it \b Estimate of the ${\cal O}(\alpha_s^3)$ and higher order corrections}\\
 Observing that the coefficients of the PT corrections for the moment decrease when $n$ increases and do not flip sign compared with the lowest moments (see Table \ref{tab:ope}) and assuming that the ratio of the 3-loop over the 4-loop coefficients are approximatively the same for each moments, we can 
write to 4-loops:
\bea
{\cal M}_{13}(8m_c^2)&\sim& 1+0.78a_s+3.06a_s^2-5.6a_s^3\nnb\\
{\cal M}_{14}(8m_c^2)&\sim& 1+0.41a_s+1.91a_s^2-3.5a_s^3\nnb\\
{\cal M}_{15}(8m_c^2)&\sim& 1+0.03a_s+0.77a_s^2-1.4a_s^3~,
\eea
where $a_s\equiv \alpha_s/\pi$. We have used the 4-loops coefficient $-5.6$ obtained in \cite{KUHN2,BOUG} for low $n=1,~Q^2=0$ moment, which we expect to be an overestimate of the coefficient of ${\cal M}_{13}(8m_c^2)$. These expressions lead to an $\alpha_s^2$ corrections of about 2\% 
for both $ r_{13/14}(8m_c^2)$ and  $r_{13/15} (8m_c^2)$ and an $\alpha_s^3$ correction of about -0.3\%
and -0.9\%. Then, taking the average of the two corrections, we may expect that the $\alpha_s^3$ corrections can provide a maximal shift of the charm quark mass of about -0.3\% leading to :
\beq
\delta_{m_c}\vert_{4-loops}\simeq \pm (2\times 4)~{\rm  MeV}~,
\eeq
a range of values expected from some alternative estimates \cite{KUHN1}.  The factor 2 in front assumes the estimate of higher order PT $({\cal O}(\alpha_s^n):~n\geq 4)$ contributions or by duality the $1/s$ corrections due to the tachyonic gluon mass $\lambda^2$ \cite{CNZ}. \\
\\
 {\it \b Final value of $\overline{m}_c$ to order ${\cal O}(\alpha_s^3)$}\\
Adding the previous estimates of new sources of contributions and errors into the 3-loops result in Eq. \ref{eq:mc3loop1}, we obtain to  ${\cal O}(\alpha_s^3)$ :
\beq
\overline{m}_c(m_c)\vert_{4-loops}=1261(18)~{\rm MeV}~.
\label{eq:mc4loop}
\eeq
This result is comparable with the existing ones obtained from moment sum rules in the literature \cite{SVZ,SNB1,SNB2,SNB3,IOFFE,KUHN1,RRY,SNmc,JAMIN,PDG}.
Our final result confirms and improves (reduction of errors) earlier sum rules analysis obtained using PT lower orders charmonium sum rules \cite{SVZ,RRY,SNmc} \footnote{More complete references can be found in Table 53.5 page 602  of  \cite{SNB1}.}. It is in agreement  with the most recent results from sum rules \cite{IOFFE,KUHN1} mentioned previously  \footnote{However, one should mention that a sum rule analysis of the $D$ meson mass  using the pseudoscalar correlator to order $\alpha_s^2$ leads to \cite{SNpseudo} $\overline{m}_c(m_c)\vert_{pseudo}=1.10(4)~{\rm GeV}$ but the value of $f_{D_s}$ agrees with the present lattice calculations. We shall reconsider this point elsewhere.}, with the lattice determination 1268(9)~{\rm MeV}   \cite{LATTmc} and with the PDG 08 average \cite{PDG}
$
\ga 1.27^{+0.07}_{-0.11}\dr~{\rm GeV}.
$
\section{ Determination of  \boldmath$\overline{m}_b(m_b)$}
\vspace*{-0.25cm}
 \nin
 We extend the previous analysis of the charmonium system to bottomium. In the following, we shall use 
 the value:
 \beq
 \alpha_s(m_b)\vert_{n_f=5}=0.219(4)~,
 \eeq 
 deduced from $\alpha_s(m_\tau)$ in Eq. \ref{eq:alphas}.  
{\scriptsize
\begin{table}[hbt]
\setlength{\tabcolsep}{1.6pc}
 \caption{\scriptsize    Masses and electronic widths of the  $\Upsilon$ family from PDG 08\cite{PDG}. }
{\small
\begin{tabular}{lll}
&\\
\hline
Name&Mass [MeV]&$\Gamma_{\Upsilon\to e^+e^-}$ [keV] \\
\hline
\\
$\Upsilon(1S)$&9460.30(26)&1.340(18)\\
$\Upsilon(2S)$&10023.26(31)&0.612(11)\\
$\Upsilon(3S)$&10355.2(5)&0.443(8)\\
$\Upsilon(4S)$&10579.4(1.2)&0.272(29)\\
$\Upsilon(10860)$&10865(8)&0.31(7)\\
$\Upsilon(11020)$&11019(8)&0.13(3)\\
\\
\hline
\end{tabular}
}
\label{tab:upsilon}
\end{table}
}
\nin
We shall use as experimental inputs the $\Upsilon$-family parameters in Table \ref{tab:upsilon} using NWA and parametrize the 
spectral function above $\sqrt{t}=(11.098\pm 0.079)$ GeV by its pQCD expression (QCD continuum), where the error in the continuum  threshold is given by the total width of the $\Upsilon(11020)$.
Using the
previous moments, the dominant contributions will come from the two lowest ground states while 
finite width corrections will not be observable. We show in Table \ref{tab:resb} the results from different
moments known to 3-loops.
{\scriptsize
\begin{table}[hbt]
\setlength{\tabcolsep}{3pc}
 \caption{\scriptsize     Value of $\overline{m}_b(m_b)$ from bottomiun moments known to 3-loops. The errors on $m_b$ come respectively from the choice of the moments, $\alpha_s$, the data on the $\Upsilon$ family and the choice of the QCD continuum threshold. The ones due to the gluon condensates are negligible here.}
{\small
\begin{tabular}{ll}
&\\
\hline
Mom&$m_b(m_b)$ [MeV] \\
\hline
\\
$Q^2$=0:\\
$r_{2/3},~r_{2/4}$&$4160(4)(2)(3)(3)$\\
\\
$Q^2$=$4m_b^2:$ \\
$~r_{8/9},~r_{8/10}$&$4177(2)(3)(3)(6)$\\
\\
$Q^2$=$8m_b^2:$\\
$r_{13/14},~r_{13/15}$&$4183(2)(4)(2)(6)$\\
\\
\hline
Average &4173(4)\\
\hline
\end{tabular}
}
\label{tab:resb}
\end{table}
}
\vspace*{-0.2cm}
\nin
\\
\\
{\it \b Error due to the subtraction point}\\
We study the effect of the subtraction point by taking $
 0.5\leq \nu^2/m_b^2\leq 2~$ and using the expression in Eq. \ref{eq:sub}. We induce an error:
 \beq
 \delta_{m_b}\vert_{\nu}=\pm 6~{\rm MeV}~.
 \eeq
 This error can be further reduced by using the 4-loops expression of the moments. \\
 \\
{\it \b Shift due to Coulombic corrections} \\
 Using $\overline{m}_b(m_b)\simeq 4.24$ GeV into  Eq. \ref{eq:coulomb}, one obtains:
\beq
 r_{Coul}\simeq 1.6 ~{\rm GeV}^{-1}~,
\eeq
 which is still larger than the confinement radius $r_{conf}\approx$ 1 GeV$^{-1}$.  These corrections can
 be render much smaller by working with a $Q^2=4m_b^2$ and large $n$ moments rather than with a $Q^2=0$ one, where the $b$-quark velocity is about 0.45 from Eq. \ref{eq:velocity}.
This value is still inside the relativistic region, where one can safely neglect these Coulombic corrections. 
 Indeed, using the previous expression of the Coulombic corrections in Eq. \ref{eq:coulomb},
we obtain the shift:
 \beq
 \delta_{m_b}\vert_{Coul}\simeq \pm 6~{\rm MeV}~,
 \eeq
 which is about the same value as the one obtained in \cite{JAMIN}. Due to the theoretical uncertainties on the real effect of the Coulombic corrections, we
consider our previous estimate as another source of errors but not as a safe correction.\\
 \\
{\it \b Value of $\overline{m}_b(m_b)$ to ${\cal O}(\alpha_s^3)$}\\
In the case of the $b$ quark, our previous estimate of the 4-loops (${\cal O}(\alpha_s^3)$) contribution  induces an error:
\beq
\delta_{m_b}\vert_{4-loops}\simeq \pm (2\times 2)~{\rm  MeV}~,
\eeq
where the factor 2 is assumed to include higher order or/and $\lambda^2$-tachyonic gluon mass corrections.
Adding these new corrections to the one in Table \ref{tab:resb},  we deduce to 4-loops accuracy:
\beq
\overline{m}_b(m_b)\vert_{4-loops}=4173(10)~{\rm MeV}~,
\label{eq:mb4loop}
\eeq
which is relatively more precise than that of $m_c$ as the non perturbative contributions are much
smaller here, while $\alpha_s$ is evaluated at a higher scale.  This  result is  
in excellent agreement with the one:
\bea
\overline{m}_b(m_b)&=&4171(14)~{\rm MeV}~,
\eea
obtained in \cite{SNcb2} using ratios of moments based on the criteria of stabilities versus
the degree $n$ (number of $Q^2$-derivatives) of moments and including $\alpha_s^3$ and the
dimension eight condensates contributions.
This  result is  also
in good agreement with
 with the PDG average \cite{PDG} :
\beq
\overline{m}_b(m_b)\vert_{PDG}=\ga 4.20^{+0.17}_{-0.07}\dr~{\rm GeV}~,
\eeq
but more precise. It also agrees with some previous results quoted in Table 53.6 (page 603) of  \cite{SNB1}.  However, it is worth mentionning that like in the case of the $D$-meson, the analysis of the B meson mass from the pseudoscalar sum rule to order $\alpha_s^2$ leads to a lower value of $(4.05\pm 0.06)$ GeV \cite{SNpseudo} which will be reconsidered elsewhere. 
\section{ Running  light and heavy quark masses at $M_Z$}
\vspace*{-0.25cm}
 \nin
For direct uses in some phenomenological applications and as inputs in some Grand Unified Model Buildings, it can be useful to convert
 these running masses $\overline{m}_Q(m_Q)$ 
 to the ones evaluated at the $Z$-mass. This can be easily done after taking care on different quark threshold effects. Using, e.g., the Mathematica  RunDec package\,\cite{STEIN}, we deduce, to 4-loops accuracy, from  Eqs. \ref{eq:mc4loop} and  \ref{eq:mb4loop}, 
the running masses evaluated at $M_Z$ for 5 flavours:
 \bea
 \overline{m}_c(M_Z)&=&616(9)_{m_c}(6)_{\alpha_s}~{\rm MeV}~,\nnb\\
 \overline{m}_b(M_Z)&=&2920(7)_{m_b}(22)_{\alpha_s}~{\rm MeV}~.
 \eea
The errors are due respectively to the values of the running mass and of $\alpha_s$ when one performs the QCD evolutions.\\
In a similar way, we can also deduce the ones of $ \overline{m}_{u,d,s}(M_Z)$ and $ \overline{m}_t(M_Z)$ by using respectively the average value from QSSR predictions to 4-loops\cite{SNmq,SNmq2}:
\bea
 \overline{m}_s(2)&=&96.1(4.8)~{\rm MeV}~,\nnb\\
 \overline{m}_d(2)&=&5.1(2)~{\rm MeV}~,\nnb\\
 \overline{m}_u(2)&=&2.8(2)~{\rm MeV}~,
 \eea
 and the on-shell top quark mass average to 3-loops \cite{PDG}:
 \beq
 M_t=171.2(2.1)~{\rm GeV}~.
 \eeq
 We obtain for 5 flavours:
 \bea
 \overline{m}_s(M_Z)&=&53.9(2.9)_{m_s}(1.9)_{\alpha_s}~{\rm MeV}~,\nnb\\
 \overline{m}_d(M_Z)&=&2.47(10)_{m_d}(3)_{\alpha_s}~{\rm MeV}~,\nnb\\
 \overline{m}_u(M_Z)&=&1.30(10)_{m_u}(3)_{\alpha_s}~{\rm MeV}~,
 \eea
 and:
 \bea
  \overline{m}_t(M_Z)&=&168.4(2.1)_{m_t}(0.1)_{\alpha_s}~{\rm GeV}~.
 \eea
 Combining the previous results, we obtain the ratios of running masses at $M_Z$:
 \bea
{{m}_b\over  {m}_c}=4.7(1)~,
\eea
and:
\bea
{{m}_s\over  {m}_u}=42(5)~,~~~~ {{m}_b\over {m}_s}=54(4)~,~~~ ~
   {{m}_t\over  {m}_b}=58(2)~,
   \label{eq:massratio}
   \eea
   where it is remarkable to notice that the ratios ${m}_s/ {m}_u$, ${m}_b/ {m}_s$ and ${{m}_t/  {m}_b}$ are almost equal which might reveal some eventual underlying
  novel symmetry of the quark mass matrix \cite{FRITZSCH,CECILIA} in some Grand Unified Theories \cite{GEORGI}. One should also observe that when one runs the ratio $m_s/m_q$ from 2 GeV to $M_Z$, the central value is not strictly constant (contrary to what expected from its renormalization group invariance) though the two values agree within the errors.  This is due to  different threshold effects and to the truncation of the series at a given order of PT. 
\section{ Conclusions}
\vspace*{-0.25cm}
 \nin
 -- Firstly, our analysis has been motivated to extract (for the first time) from the sum rules, the instanton liquid model radius $\rho_c$ in Eq. \ref{eq:rho} , which parametrizes the ratio of the $\la g^3f_{abc}G^3\ra$ over the $\la \alpha_s G^2\ra$ gluon condensates, where the corresponding value of $\la g^3f_{abc}G^3\ra$ in Eq. \ref{eq:gg3} is much larger than usually quoted in the literature.  However, despite this large value, the OPE in the ratios of moments which we have used continues to present a good convergence. Because we have neglected the contributions of higher dimension condensates in our approach, one may consider this value of $\la g^3f_{abc}G^3\ra$ as that of an {\it ``effective gluon condensate"} which may include in it all the higher dimension condensates contributing to the OPE and not considered in our analysis.\\
-- Using the previous result and thanks to the recent progresses in evaluating accurately the pQCD series of the heavy quarks vector correlators \cite{KUHN,KUHN2,BOUG} and to more accurate measurements of the corresponding spectral functions \cite{PDG,BES,BABAR}, it becomes possible to extract with a high precision the heavy quark masses using higher $n$ ratios of moment sum rules. The results for $m_c$ and $m_b$ in Eqs. \ref{eq:mc4loop} and  \ref{eq:mb4loop}, where different sources of errors are under a good control (see also the comments in section 5), are among the most accurate measurements available today. 
These results confirm and improve estimates done in the early days of sum rules  \cite{SVZ,RRY,SNmc,SNB1,SNB2,SNB3}\,\footnote{One can notice, in different papers written by the author for extracting  $m_c$ and $m_b$ from heavy quarkonia sum rules, that their central values remain very stable since the 1st paper in 1987 \cite{SNmc}. This feature indicates (a posteriori) the self-consistency of the approach and the good convergence of the PT and OPE, especially when one works with the running ${\overline{MS}}$ masses.}. \\
-- Compared with some other recent determinations based on low-$n$ and $Q^2=0$ moments \cite{KUHN,KUHN1,KUHN2,BOUG}\,\footnote{The experimental results in the high-energy regions \cite{PDG} do not strictly co\"\i ncide with the pQCD predictions. In this approach a shift of about 5\% in the continuum would affect by 64, 21 and 9 MeV the value of $m_b$ from ${\cal M}_{1,2,3}(0)$ moments.}, our approach should (a priori) be more accurate because we work with ratios of moments which are less sensitive to the continuum contribution than the individual moments. We also refrain to take too high moments where their QCD expressions can become difficult to control. The apparent accuracy of the  results obtained in the current literature are also due (among others)  to the neglect of the gluon condensates (see Fig. 2) which are one of the main sources of the errors in the determinations of $m_c$ (and of $\alpha_s$) from moment sum rules (see Table \ref{tab:res}). The same remarks also apply to the higher-$n$ moments used in \cite{JAMIN,IOFFE}. \\
-- The agreement of the present results with the most precise recent lattice calculation of $m_c$ \cite{LATTmc} confirms the robustness of the higher $n$ ratios of moments sum rule approach and, in general, the ability of QSSR to extract reliably the QCD parameters from hadron properties. It becomes now challenging  to check our estimate of $m_b$ given in Eq. \ref{eq:mb4loop} using lattice calculations.\\
-- Finally, the approximate equalities of the different ratios of the quark masses in Eq. \ref{eq:massratio}, when they are evaluated at $M_Z$ with the same number of flavours might reveal some eventual underlying  novel symmetry  of the quark mass matrix in some Grand Unified Theories.
\section*{Acknowledgements} 
\vspace*{-0.25cm}
\nin
I wish to thank Valya Zakharov for reading the preliminary draft and for some comments and Adriano Di Giacomo for some communications on the lattice results of the gluon condensates. I also thank the 2nd reviewer for a careful reading of the manuscript and for some constructive comments leading to an improvement of the paper.
\vspace*{-0.25cm}
\input{bib_charm2}

\end{document}